\documentclass[sigconf,natbib=false]{acmart}

\AtBeginDocument{%
  }

\usepackage{amsmath,amsfonts}
\usepackage{algpseudocode}

\DeclareMathOperator*{\argmax}{arg\,max}
\usepackage[ruled, vlined, norelsize, linesnumbered]{algorithm2e}
\usepackage{algorithmicx}
\usepackage{graphicx}
\usepackage{textcomp}
\usepackage{mathtools}
\usepackage{xcolor}
\usepackage{booktabs}
\SetKwInput{KwInput}{Input}
\SetKwInput{KwOutput}{Output}

\usepackage{acronym}
\usepackage{enumitem}

\acrodef{AIG}{and-inverted graph}
\acrodef{DAG}{directed acyclic graph}
\acrodef{CPA}{correlation power attack}
\acrodef{QoR}{quality of result}
\acrodef{PCC}{Pearson correlation coefficient}
\acrodef{PSC}{Power side-channel}
\acrodef{PSCA}{power side-channel attack}
\acrodef{RTL}{register transfer level}
\acrodef{SA}{simulated annealing}
\acrodef{PPA}{power, performance, and area}
\acrodef{MCTS}{Monte-Carlo tree search}
\acrodef{PT}{power trace}
\acrodef{MDP}{Markov decision process}
\acrodef{RL}{reinforcement learning}
\acrodef{UCT}{upper confidence tree}
\acrodef{LVT}{low voltage threshold}
\acrodef{HVT}{high voltage threshold}
\acrodef{FPGA}{field programmable gate array}
\acrodef{CLB}{configurable logic block}
\acrodef{SOTA}{state-of-the-art}
\usepackage{pifont}
\usepackage{soul}

\usepackage{multirow}
\usepackage[caption=false,font=footnotesize,labelformat=simple]{subfig} 
\usepackage{acronym}
\newcommand{\solution}{ASCENT}

\usepackage{balance}
\newcommand{\circleone}{\ding{202}}
\newcommand{\circletwo}{\ding{203}}
\newcommand{\circlethree}{\ding{204}}

\acmConference[Conference acronym 'XX]{Make sure to enter the correct
  conference title from your rights confirmation emai}{June 03--05,
  2018}{Woodstock, NY}
\acmISBN{978-1-4503-XXXX-X/18/06}

\RequirePackage[
  datamodel=acmdatamodel,
  style=acmnumeric,
  ]{biblatex}

\begin{document}

\title{\solution{}: \underline{A}mplifying Power \underline{S}ide-\underline{C}hannel R\underline{e}silience via Lear\underline{n}ing \& Monte-Carlo \underline{T}ree Search}

\author[Bhandari, Chowdhury, et al.]{
Jitendra Bhandari$^1$\textsuperscript{*}, Animesh Basak Chowdhury$^1$\textsuperscript{*},  Mohammed Nabeel$^2$, 
Ozgur Sinanoglu$^2$, Siddharth Garg$^1$,
Ramesh Karri$^1$,
Johann Knechtel$^2$}
\affiliation{\institution{$^1$New York University, USA, $^2$New York University Abu Dhabi, UAE}
\country{}
}

\begin{abstract}
Power side-channel (PSC) analysis is pivotal for securing cryptographic
hardware.
Prior art focused on securing
gate-level netlists obtained as-is from chip design automation, neglecting all the
complexities and potential side-effects for security arising from the design automation process. That is, 
automation traditionally prioritizes power, performance, and area (PPA), sidelining security.
We propose a ``security-first'' approach, refining the logic synthesis stage to
enhance the overall resilience of PSC countermeasures. We
introduce ASCENT, a learning-and-search-based framework that (i)~drastically reduces
the time for post-design PSC evaluation and (ii)~ explores the security-vs-PPA design space.
Thus, ASCENT enables an efficient exploration of a large number of candidate
netlists, leading to an improvement in PSC resilience
compared to regular PPA-optimized netlists.  ASCENT is up to \texttt{120x}
faster than traditional PSC analysis and yields a \texttt{3.11x}
improvement for PSC resilience of state-of-the-art PSC countermeasures.
\end{abstract}


\setcopyright{acmlicensed}
\copyrightyear{2024}
\acmYear{2024}
\acmDOI{XXXXXXX.XXXXXXX}

\settopmatter{printacmref=false, printccs=true, printfolios=true}


\maketitle

{
\renewcommand{\thefootnote}{\fnsymbol{footnote}}
\footnotetext[1]{J. Bhandari and A. B. Chowdhury contributed equally to this work.
}}


\section{Introduction}
\label{sec:intro}


Hardware implementations of cryptographic and other sensitive algorithms are well-known to be vulnerable to side-channel attacks.
Kocher \textit{et.} al~\cite{kocher} first demonstrated the \ac{PSCA} by exploiting variations in power profiles to extract the secret keys; various advanced PSCA versions followed throughout the years~\cite{survey-PSC}.
To counter such attacks, there are many ongoing efforts to enhance the
\ac{PSCA} resilience of hardware implementations.
For example, the \ac{SOTA} countermeasures~\cite{mask-2005,Moos_Moradi_2021, 7324539}
augment secret data with random noise to obscure power profiles from secret data. However, all these countermeasures incur large \ac{PPA} overheads.

\begin{figure}[t]
\centering
\includegraphics[width=\columnwidth]{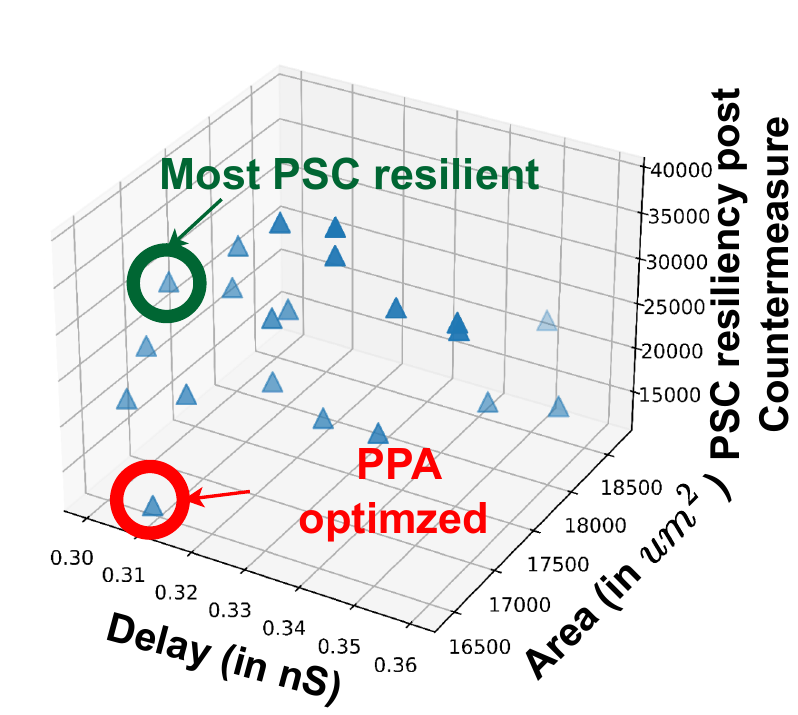}
\caption{PSCA resilience post-integration of the \textit{QuadSeal} countermeasure~\cite{7324539}
	vs. area-delay of various AES netlists.
}
    \label{fig:introPlot}
\end{figure}

To tackle PPA overheads, security researchers typically take the outputs of chip design automation processes---which are optimized for PPA by default---as a starting point, perform PSCA analysis, and then
propose/apply countermeasures to mitigate PSCAs.
However, such an approach can easily overlook circuit configurations that might be inherently more
more effective to support the resilience of PSCA countermeasures, even if they have some PPA disadvantages.
We demonstrate this in \autoref{fig:introPlot}, where we show
the area-delay plot of various synthesized netlists versus the final PSCA resilience,
for a representative AES hardware implementation with the \textit{QuadSeal} countermeasure \cite{7324539} applied (see also Sec.~\ref{sec:counter} for the latter).


Our findings in \autoref{fig:introPlot}
clearly demonstrates that \ul{optimizing the baseline netlist for PPA alone does not guarantee the strongest possible defense against PSCAs in the end}. This highlights the need to explore
alternative circuit configurations for inherently better PSCA
resilience, leading 
to our core research questions:
\begin{enumerate}[leftmargin=*]
    \item What is the right circuit implementation to start with, such that---post-countermeasure application---the hardware would have the best \ac{PSCA} resilience?
    \item How can we guide chip design automation to generate such highly PSCA-resilient circuits, yet with low PPA overheads?
\end{enumerate}


In this work, we systematically tackle the fact that different logic synthesis approaches can significantly impact PSCA resilience (for better or worse).
This is because different optimization steps
result in different netlists with varying characteristics for the type, number, driver strengths, etc. of the standard cells used.
Naturally, all these directly impact the power profiles, thereby 
impacting the prospects of PSCAs as well.  However, the related design-space exploration is a tedious, parameter-rich challenge. Furthermore, with commercial
synthesis tools, we face a black-box optimization problem, where the
relationship between synthesis choices and PSC resilience is difficult to model.

\begin{figure}[t]
\centering
\includegraphics[width=0.8\columnwidth]{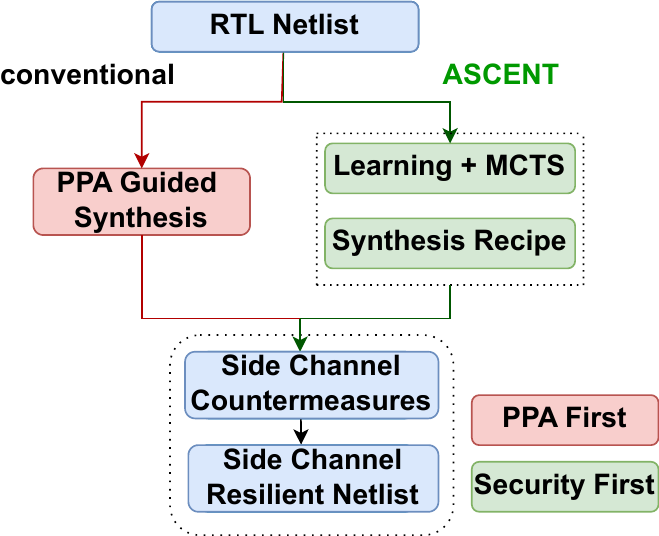}
\caption{High-level view on ASCENT framework for optimized PSC resilience of countermeasures.}
    \label{fig:introDiag}
\end{figure}

To address these challenges,
we propose ASCENT, a framework for \textit{amplifying PSC resilience via learning and \ac{MCTS}}
	 (\autoref{fig:introDiag}).
Our key contributions are:
\begin{itemize}[leftmargin=*]
    \item
    ASCENT provides a hybrid learning-and-search approach, enabling efficient and effective exploration of the security-centric design space.
    ASCENT enables us to explore 120$\times$ more configurations compared to a naive search.
    \item ASCENT helps us to achieve up to 3.11$\times$ improvements in PSC resilience for \ac{SOTA} countermeasure integration when compared to regular, PPA-optimized baseline netlists. At the same time, the PPA impact is well controlled and limited by ASCENT, 
    namely only up to 6.61\% more area.
    \item We open-source ASCENT as a commitment towards reproducible research.
    All the algorithms, experiments, and benchmarks are publicly available at~\url{https://github.com/NYU-MLDA/scarl.git}.
\end{itemize}

\section{Background and Motivation}
\label{sec:background}



\subsection{Power Side-Channel Attacks (PSCAs)}
\label{sec:bg:PSCAs}

Cryptographic hardware is vulnerable to PSCAs, which exploit the fluctuations in a device's power consumption to reveal sensitive information like 
keys \cite{survey-PSC}. Thus, these attacks leverage the fundamental connection between a device's power consumption and its internal state during operations.
There are different types of PSCAs, including simple power analysis (SPA), differential power analysis (DPA), and \ac{CPA} \cite{brier2004CPA}.\footnote{%
	SPA directly interprets the power profiles during specific cryptographic operations, aiming to deduce sensitive data.
	DPA goes further by comparing power consumption across multiple similar operations with varying inputs.
	This technique seeks to isolate variations in power profiles that directly correlate with the influence of the secret key.
	CPA employs statistical tools, most commonly the \ac{PCC}, to match hypothesized power consumption patterns
	against the actual power
	measurements. The highest correlation often reveals the correct key.
	Note that we utilize CPA in this work; more details are provided further below.}


\textbf{Technology Implications.}
With continuous advancements for the ever-shrinking technology nodes, the threat of static power side-channel attacks (S-PSCAs) has significantly increased over the years \cite{leakage-2010-TCAS,amstatic2014,giorgetti2007}.
Unlike dynamic PSCAs, which focus on power fluctuations during active computations, S-PSCAs exploit the relationship between stored data and static power consumption (leakage power).
More specifically, advanced nodes utilize standard cells of various types with different power-performance characteristics,
e.g., low-threshold voltage (LVT) and ultra-low threshold voltage (ULVT) cells are faster than the regular (RVT) cells, thereby helping to meet faster timing constraints, albeit at the expense of significantly higher leakage power.
LVT and ULVT cells are essential for timing closure, i.e., the careful final-stage efforts in design automation.
In short, these implications highlight the need for dedicated countermeasures that protect against S-PSCAs. 




\textbf{Countermeasures.}
To combat S-PSCAs, security-aware designers can employ masking \cite{mask-2005,bhandari2024lightweight}, shuffling, and/or balancing \cite{Moos_Moradi_2021, 7324539} schemes.\footnote{%
Further countermeasures and details are discussed in Section~\ref{sec:bg:spsc}.
Also note that the specific countermeasures employed for this work are discussed in Section~\ref{sec:counter}.}
In general, these strategies seek to obscure the power profiles.
Despite their demonstrated effectiveness, they all increase design overheads considerably, necessitating to strive a careful balance between security and PPA during the design process.

As indicated,
here we tackle this challenge through our novel, learning-and-search based framework for logic synthesis.

\textbf{{Simulation-Based Power Analysis.}}
This is crucial for understanding a design's vulnerability to PSCAs before investing into actual tape-outs.
Commonly utilized procedures work along the following lines; we employ such a procedure as well in this work.

First, through gate-level simulations, a value change dump (VCD) file is obtained. This captures all the relevant state information of the device under test.  In combination with the post-synthesis netlist and the library files,
	this allows for accurate power analysis.
Second, power simulation tools calculate the power consumption of each cell, including static/leakage power and internal power from input/output switching transitions.  Importantly, for S-PSCA
assessments, zero-delay simulations enable precise static power capture during specific operations (unlike an averaged leakage power analysis provided by full-timing simulations).


\textbf{{Correlation Power Analysis (CPA).}}
This attack hinges on identifying a correlation between a device's power consumption and the intermediate data processed throughout cryptographic operations.  An attacker collects power consumption data and then hypothesizes
on all possible intermediate data values, which often involve direct correlations to parts or derivatives of the secret key.

More specifically, power consumption is predicted for each hypothetical intermediate value,
typically using models like Hamming weight or Hamming distance, which relate
binary data representation to power. The core of CPA involves calculating the
PCC between actual power measurements and the predictions for each hypothesis.
The hypothesis yielding the highest correlation often represents the correct
assignment for some part/derivative of the key, allowing the attacker to
reconstruct the full key eventually.

\subsection{Related Work}


\textbf{S-PSC Attacks and Countermeasures.}
\label{sec:bg:spsc}
\cite{giorgetti2007} showed, for the first time, the potential of S-PSCAs as a severe threat.
\cite{amstatic2014} have conducted one of the first practical experiments for S-PSCAs using FPGAs, with some follow-up work presented in \cite{7927198}.
\cite{leakage-2010-TCAS} highlighted the importance of leakage power and its effect on PSCAs especially for advanced nodes.
\cite{Moos_Moradi_2021} proposed various countermeasures against S-PSCAs, albeit with considerable PPA overheads.
\cite{bhandari2023lightweight} have shown the impact of various types of standard cells on S-PSCAs.
\cite{bhandari2024lightweight} proposed a lightweight masking scheme against S-PSCAs.
\cite{Karimi_Moos_Moradi_2019} have demonstrated the important side-effect of aging for S-PSCAs in advanced technology nodes.
%
\cite{10.1007/978-3-319-57339-7_5} studied multivariate techniques focused on leakage power consumption to enhance cryptographic security assessments.
\cite{9040870,cryptography5030016} proposed standard-cell, delay-based dual-rail pre-charge logic (SC-DDPL) as countermeasure.
However, due to its structural complexity, this countermeasure is incompatible with commercial design flows.

\textbf{Design Frameworks for Advancing PSC Resilience.}
\cite{karna} introduced a framework that scores and optimizes design parts to minimize PSC vulnerabilities. This approach is limited by the impractical assumption of
timing slacks being ubiquitously available. It also lacks an actual PSC evaluation.
%
\cite{rtl_psc} proposed a framework for assessing PSC vulnerabilities at the register-transfer level (RTL), with the goal to aid countermeasure implementation.
\cite{7364404} studied circuit replication and SRAM sharing for PSC resilience in FPGAs. This approach notably increases design costs but significantly improves security while maintaining FPGA configurability.
\cite{10.1145/3488932.3517415} emphasized the importance of automated modeling for early, system-level detection of potential leaks.
%
%
\cite{Tiri2004SecureLS} proposed a methodology for so-called wave dynamic differential logic (WDDL) on FPGAs.
%

\textbf{Summary.}
Prior art for S-PSC countermeasures is focused on detailed empirical studies, with only limited considerations for generalized design-time integration of the countermeasures.
At the same time, prior art for frameworks is limited to D-PSCAs, not S-PSCAs, and was proposed for high-level design stages and/or for FPGAs.
Thus, \ul{there is no prior art that proposes a security-first approach toward the complex, yet critical, challenge of design-space exploration for S-PSCA countermeasure integration in ASICs.} Also recall the exploratory finding from Section~\ref{sec:intro}.
This gap
provides the main motivation for our work.


\subsection{Representative
Countermeasures}
\label{sec:counter}


As motivated in Sections~\ref{sec:bg:PSCAs} and \ref{sec:bg:spsc}, S-PSCAs are becoming ever-more relevant for advanced technology nodes, and various countermeasures have been proposed.
In our study, we consider the following two representative, \ac{SOTA} countermeasures against S-PSCAs.\footnote{%
While these two SOTA countermeasures appear similar from a high-level view, their implementation details and, thus, efficiency to hinder S-PSCAs still differ \cite[Table~5]{Moos_Moradi_2021}.
} \ul{Importantly, ASCENT is agnostic to the countermeasures a designer likes to explore and eventually integrate.}
	

\textbf{{Quadruple Algorithmic Symmetrizing (QuadSeal)~\cite{7324539}.}} This technique can protect against both dynamic and static PSCAs.
It focuses on achieving a balance in Hamming weights/distances for the cryptographic operations in hardware.
It operates by quadrupling the unprotected circuit structure and balancing the arrangement of the so-called \textit{substitution boxes (S-Boxes)}
in three of those circuit copies in a specific manner.
Additionally, it involves rotating inputs to the resulting balanced structure, to mitigate other real-world dependencies introduced by, e.g., manufacturing process variations, timing-path imbalances, aging, etc.

\textbf{{Exhaustive Logic Balancing (ELB)~\cite{Moos_Moradi_2021}.}}
To reduce the correlation between input data and the leakage current of standard cells, selected sensitive cells are duplicated and fed with inverted input data, which is akin to differential logic.
Cell duplication is scaled based on the number of possible input vectors---a single-input cell is duplicated once, while two-input cells are quadrupled. This, along with the inverted inputs, ensures a constantly uniform
input distribution across all related cells.

\begin{figure*}[t]
    \centering
    \includegraphics[width=\textwidth]{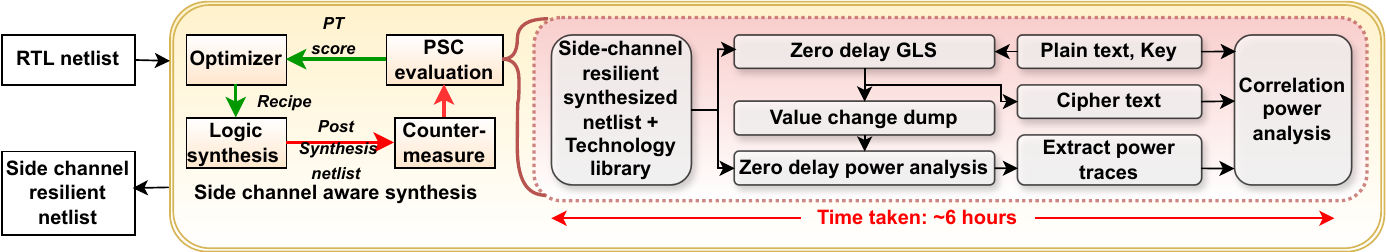}
    \caption{An exemplary PSC-aware logic synthesis framework. Despite the focus on security, conventional flows utilize PPA-optimized netlists as starting point to integrate PSC countermeasures right-away on top (red arrows), without
	    further feedback loops.
     The latter omission is due to the significant runtime taken by PSC evaluation (light-red, dotted box). Naturally, this can result in sub-optimal countermeasure integration. In contrast, our work utilizes such essential feedback (green arrows) and achieves this by an efficient learning-and-search approach (detailed in \autoref{fig:ascent-framework}).
     }
    \label{fig:problem_statment}
\end{figure*}

\subsection{Logic Synthesis}
\label{subsec:logic_synthesis}



Logic synthesis transforms a high-level hardware design (e.g., in RTL) into an optimized and technology-specific gate-level netlist. This process offers significant flexibility, as any single design can be mapped to many functionally equivalent but structurally different netlists, all with distinct PPA characteristics.
Toward that end, so-called recipes are devised, which are sequences of optimization steps.
However, the sheer number of possible recipes and netlists makes this a complex problem, in fact a
      $\Sigma^P_2$-complete problem.



\textbf{Setting.}
Commercial tools like \textit{Synopsys DC} and \textit{Cadence Genus} are tuned for PPA optimization and
use proprietary optimization algorithms toward that end.
Aside from scripting interfaces tailored for such PPA optimization, these tools lack direct mechanisms to tune synthesis for other objectives like PSC resilience.

Thus, research into novel optimization techniques, including our work on security-first synthesis, often relies on the open-source and customizable Yosys framework \cite{yosys}.
In fact, Yosys is the most widely adopted, SOTA synthesis framework for and by academia.


\textbf{AIG Representation, Generic Problem Formulation.}
Within Yosys, ABC~\cite{abc} is used for combinational optimization. First, ABC converts the design into a homogeneous logic-network implementation called \ac{AIG}.
Next, ABC's algorithms employ various transformations at the sub-graph-level~\cite{abc}. Importantly, users are free to tweak these algorithms in general and the selection and order of transformations in particular, all to optimize the AIG circuit representation according to their objectives.

More formally, in line with earlier works~\cite{chowdhury2021openabc,chowdhury2022bulls,chowdhury2024retrieval} a synthesis recipe $a^T$ is a sequence of $T$ transformation steps operating on an AIG structure to optimize for PPA and/or other metrics (e.g. security~\cite{basak2023almost}), all while preserving the original functionality.
We denote $\mathcal{A}$ of $\mathbf{L}$ unique synthesis transformations, $\{a_{0}, a_{1}, \ldots, a_{L-1}\}$ ($a_{i} \in \mathcal{A}$), in a synthesis recipe $a^T$.
Thus, the number of  synthesis recipes of length $T$ is  $\mathbf{L}^\mathbf{T}$, including repeatable transformations. This search space is denoted by $\mathcal{A}^T$.
The problem of generating a PPA-optimal synthesis recipe for an AIG is:
\begin{align}
\argmax_{a^T \in \mathcal{A}^T} PPA(AIG_T),\ \ s.t. \, \, AIG_{t+1} = \eta(AIG_{t},a_{t}) \, 
\forall t\in [0,T-1]
\end{align}
where $\eta$ is the synthesis function defined as $\eta : AIG \times \mathcal{A} \longrightarrow AIG$.

\subsection{Monte-Carlo Tree Search}
\label{sec:bg:MCTS}

\ac{MCTS} is an optimization algorithm best suited for tree-structured search-space exploration.
It has been used in selected prior art for logic synthesis~\cite{yu2020flowtune,pei2023alphasyn,chowdhury2024retrieval,delorenzo2024make}, albeit only for \ac{PPA} optimization.

\textbf{Structure.}
The MCTS search tree contains a root node representing the initial state ($S_0$).
A node is called leaf if there exist an $a_t \in \mathcal{A}$ that still remains unexplored and the node is terminal state.
Each node preserves two attributes: (1)~node visit count $N(S_t,a_t)$ and (2)~cumulative reward $R(S_t,a_t)$.\footnote{%
$N(S_t,a_t)$ is the number of times the nodes is visited during exploration.
$R(S_t,a_t)$ is the total reward obtained while exploring the sub-tree rooted at that node.}

MCTS operates in four stages as follows.

\textbf{1) Selection:}
of the ``most promising node'' in the MCTS tree until a leaf node is reached for further exploration.
The selection is based on the \ac{UCT} computation as follows:
\begin{equation}
\pi_{MCTS}(S_t) = \argmax_{a_t \in \mathcal{A}} \left(\underbrace{\frac{R(S_t,a_t)}{N(S_t,a_t)}}_{\text{Exploitation}} + c\dot{\underbrace{\sqrt{\frac{\log\sum_{a_t\in\mathcal{A}}N(S_t)}{N(S_t,a_t)}}}_{\text{Exploration}}}\right)
\label{eq:mcts}
\end{equation}
That is, \ac{UCT} computation considers \textit{exploitation} and \textit{exploration}.

\textit{Exploitation} computes the ratio of the reward $R(S_t,a_t)$
accumulated over the sub-tree rooted at that node and the visits count $N(S_t,a_t)$.
This average reward is obtained by exploring the sub-tree which, in turn, is done by picking an action $a_t$ from state $S_t$.

\textit{Exploration} prioritizes nodes which have been less explored so far.
It loosely computes the
ratio of the parent's visit count ($\sum_{a_t\in\mathcal{A}}N(S_t,a_t)$) and the current node's visit count ($N(S_t,a_t)$).
Thus, the score increases if the parent node has been frequently visited whereas the current node has been less explored.

Starting from state $S_0$, $\pi_{MCTS}$ navigates the search space by selecting the ``best'' action $a_t$
that achieves the maximum score combining exploitation and exploration terms (\autoref{eq:mcts}), effectively performing a best-first exploration.
The selection process continues until a leaf node is encountered.

\textbf{2 and 3) Expansion and Rollout:}
Once a leaf node is selected, an action $a_t$ is chosen (at random) from the set of unexplored actions.
A node is added to the MCTS tree and node attributes are initialized.
We call the trajectory $\tau$ a
sequence of nodes visited during MCTS selection and expansion, which is represented by $\{S_0, (S_0,a_0), (S_1,a_1).., (S_{\tau},a_{\tau})\}$.

\textbf{4) Backpropagation:}
After computing the score for the terminal state $S_T$, the trajectory $\tau$ is backtracked from the leaf node till the root node.
The cumulative reward $R(S_t,a_t)$ and node visit count $N(S_t,a_t)$ for each node in
$\tau$ are updated with $R(S_T)$ and $1$, respectively.
Next, the process repeats from Stage 1) again.

\section{Problem Formulation}
\label{sec:problem}

We believe that logic synthesis offers ample opportunities to discover inherently more S-PSCA-resilient netlist structures that can amplify the resilience of SOTA countermeasure even further.
However, while traditional PPA-focused synthesis does impact S-PSCA outcomes, recall that a dedicated security-first approach is fundamentally missing (Section~\ref{sec:bg:spsc}).
We understand and emphasize that realizing such an approach requires enormous efforts in practice.
This is due to two key facts:
\begin{enumerate}[leftmargin=*]
\item synthesis in general is already complex and its search-space computationally expansive to explore (see Section~\ref{subsec:logic_synthesis});\\
	and, coming on top,
\item actual S-PSC evaluation, which is essential for accurate guidance for a security-first synthesis method, incurs significant further computation cost (see further below).
\end{enumerate}

\autoref{fig:problem_statment} illustrates this problem for an exemplary, PSC-aware synthesis framework.
Next, we formalize this problem in general.
Subsequently, we evidence the practical challenges outlined above in more detail.
We also indicate on the techniques we utilize to address these.
Finally, we provide the specific problem formulation, leading to the proposed ASCENT framework.

\begin{figure*}
    \centering
    \includegraphics[width=\textwidth]{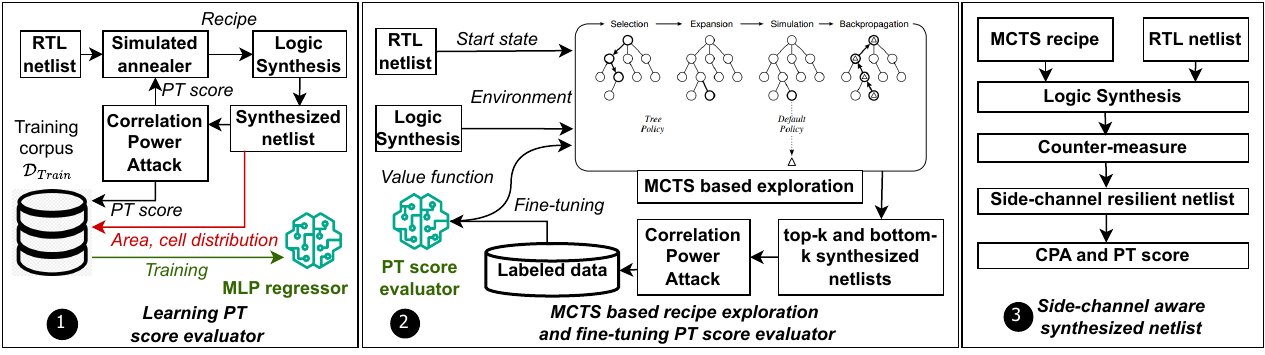}
    \caption{The ASCENT framework. \circleone{} Train a $PT_{score}$ predictor for ultra-fast \ac{PSC} evaluation of synthesized netlists. \circletwo{} MCTS-based search-space exploration to obtain a synthesis recipe that maximizes $PT_{score}$. \circlethree{} use the obtained recipe for security-first synthesis, apply the countermeasure of choice on top, and validate $PT_{score}$ of the final netlist.}
    \label{fig:ascent-framework}
\end{figure*}

\textbf{Formulation.}
Our goal is to find a netlist
that maximizes S-PSCA resilience after application of SOTA S-PSCA countermeasure.
To address these challenges, we formulate security-first synthesis as an optimization problem, guided by a \ac{MDP}, with distinct states, actions, transitions, and rewards.
%

\begin{itemize}[leftmargin=*]
\item \textit{State $S_t$} at step $t$ is the \ac{AIG} of the  design $D$ after applying a partial synthesis recipe of length $t$. $AIG_0$ is the initial AIG extracted from $D$. The terminal state $AIG_T$ is the  \ac{AIG} generated after applying a synthesis recipe of maximum length $T$.

\item \textit{Actions $\mathcal{A}$} is the set of $L$ functionality-preserving transformations $\{a_{0}, a_{1}, \ldots, a_{L-1}\}$ ($a_{i} \in \mathcal{A}$) provided by a synthesis tool.

\item \textit{State transition $\eta(S_{t+1}|S_t,a_t)$} is the transformation by applying action $a_t$ on state $S_t$ resulting in state $S_{t+1}$. 
Here, the transition function yields deterministic \ac{AIG}.

\item \textit{Reward}
expresses the S-PSCA resilience as so-called \ul{$PT_{score}$, i.e., the number of power traces required for successful key extraction,}
of the post-countermeasure netlist.
We consider a delayed
reward model and assign zero reward to every action until we reach terminal state $S_T$.
\end{itemize}
We define the overall problem as $PT_{score}$ maximization:
\begin{align}
\argmax_{a_{T} \in \mathcal{A}^T}  PT_{score}(\mathcal{C}(S_T)), \, \, s.t. \, \, S_{t+1} = \eta(S_{t},a_{t}) \, \, \forall t\in [0,T-1]
\end{align}
where $\mathcal{C}$ is the countermeasure applied on the synthesized netlist.

\textbf{Practical Challenges.}
%
As indicated, there are critical barriers in terms of computational complexity associated with this problem.
For example, the search space for synthesis in general is of complexity $L^T$, where $L=13$ and $T=18$,\footnote{%
    This exemplary choice of $L=13$ and $T=18$ is in line with the length of synthesis recipes and unique synthesis transformations available for Yosys' \texttt{compress2rs} recipe.}
    is approximately $\sim10^{19}$.

As our problem has a non-analytical form and no closed-form solution is available, we must rely on
gradient-free optimization methods. This mandates means for inexpensive reward evaluation.
However, our experimentation shows that running an accurate PSC attack evaluation on
post-countermeasure netlists requires up to 100k test vectors, which takes $\approx 6$ hours of simulation runtime. Thus, even when evaluating only 100 samples using any gradient-free
optimizer, the process would take $\approx 25$ days.

These important observations raise the following questions toward computationally-efficient S-PSC-aware logic synthesis:
\begin{itemize}[leftmargin=*]
    \item Given a runtime budget, how can we quickly, yet accurately, evaluate S-PSC attacks for some post-countermeasure netlist?
    \item How can we efficiently explore the search space of synthesis recipes to obtain S-PSC-resilient netlists?
\end{itemize}

\textbf{Outline of Method.}
Addressing these challenges necessitate an optimization approach that balances search efficiency with accurate S-PSCA assessment.
This leads us to the design of ASCENT (Section~\ref{sec:method}), which employs a hybrid learning-and-search strategy.
More specifically, ASCENT utilizes
(i)~a \textit{zero-shot predictor} $\hat{PT}_{score} (\mathcal{C}(S_T),\theta)$
to significantly speed up the PSC evaluation, without loss of accuracy, and
(ii)~MCTS (Section~\ref{sec:bg:MCTS}) to explore the large and complex search space for security-first synthesis in an effective and efficient manner.



\textbf{Extended Formulation.}
%
Assume a predictor $\hat{PT}_{score} (\mathcal{C}(S_T),\theta)$, which predicts the number of power traces required for PSCAs for a design $\mathbf{D}$ synthesized using a $T$-length recipe.
Then, we seek to solve the following problem:
\begin{align}
\label{eq:final}
\argmax_{a_{T} \in \mathcal{A}^T}  \hat{PT}_{score}(\mathcal{C}(S_T),\theta), \, \, s.t. \, \, S_{t+1} = \eta(S_{t},a_{t}) \, \, \forall t\in [0,T-1]
\end{align}
where $\mathbf{\theta}$ represents the predictor's parameters.
In simple terms, the predictor shall serve as a computationally efficient surrogate for direct S-PSCA evaluation.


\section{ASCENT Framework}
\label{sec:method}


We solve the problem formulation in Equation~\ref{eq:final} in three steps, as illustrated in \autoref{fig:ascent-framework}.
Next, we outline these three steps.

First, we provide an exploratory experiment which clearly demonstrates that maximizing the S-PSCA resilience of a netlist post-countermeasure integration, i.e., maximizing $PT_{score}(\mathcal{C}(S_T))$, is the same
as maximizing the $PT_{score}(S_T)$.
Therefore, we train a {zero-shot} $\hat{PT}_{score}$ predictor using $PT_{score}$ values obtained from diverse synthesized netlists.
We show that it suffices to start with some representative data points to train such zero-shot predictor.

Second, we employ the zero-shot predictor
as S-PSCA evaluator for the MCTS-based exploration of the search space. This predictor within MCTS is essential to drive the sequential decision-making process.
Note that we also improve the predictor on the fly:
we simulate the actual power traces for the top-$k$ and bottom-$k$ recipes during the \ac{MCTS} process and accordingly fine-tune the predictor.

Third, we put all parts together and conduct an end-to-end optimization process, including a final validation of the PSC resilience by actual PSC evaluation after the optimization process is done.



\subsection{Zero-Shot
Predictor (\circleone{})}
\label{sec:proxy}

To obtain a zero-shot predictor, we have to collect historical S-PSCA data in a one-time effort.
Key challenges here are (i)~high runtime cost of S-PSCA evaluation even for such one-time effort and (ii)~ensuring a representative dataset of diverse netlists with varying S-PSCA resilience.
Next, we discuss how we tackle these challenges and finally outline the actual training approach.


\subsubsection{Pre- vs Post-Countermeasure Evaluation}

\begin{figure}[t]
\centering
\includegraphics[width=\columnwidth]{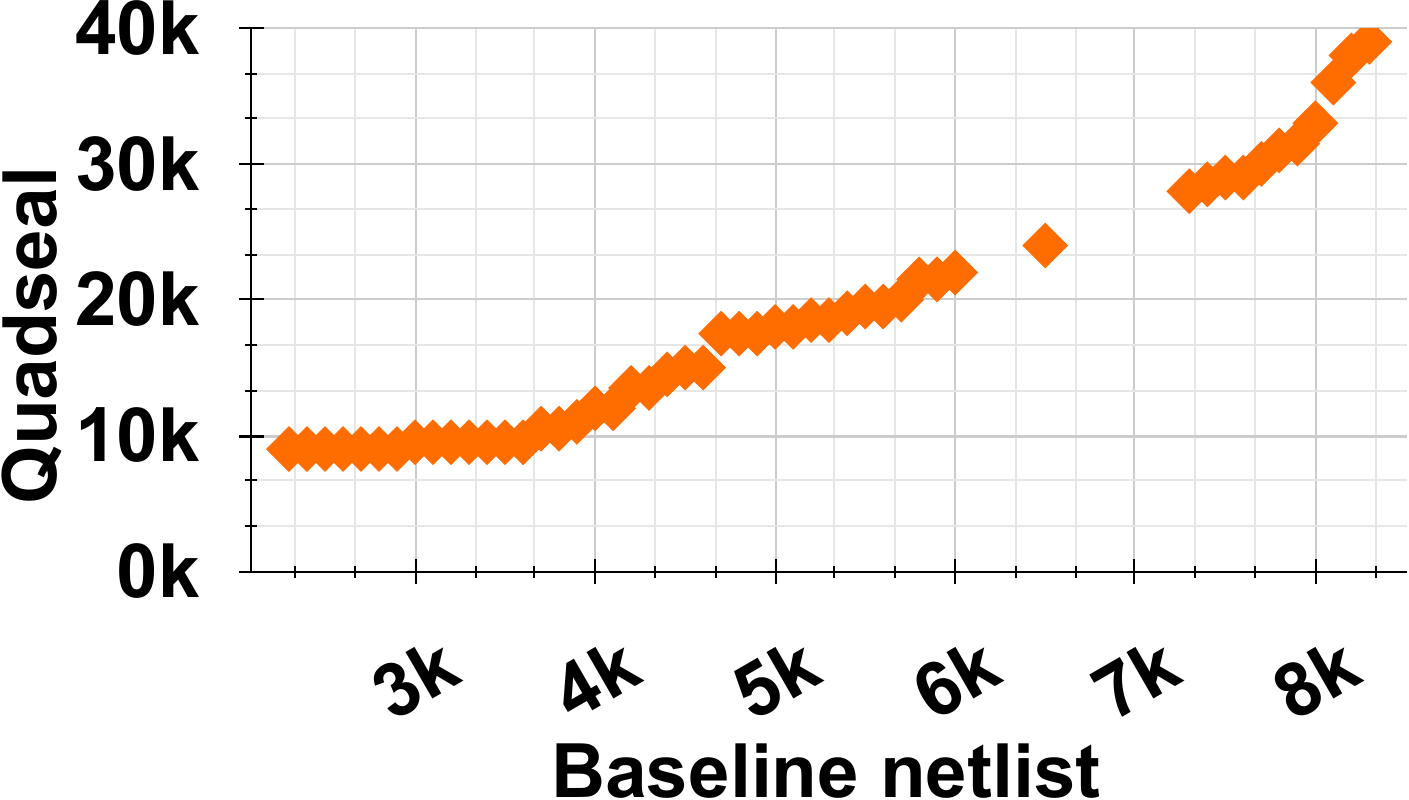}
\caption{$PT_{score}$ on AES before and after applying QuadSeal.}
    \label{fig:quadseal_motiv}
\end{figure}
\begin{figure}[t]
\centering
\includegraphics[width=\columnwidth]{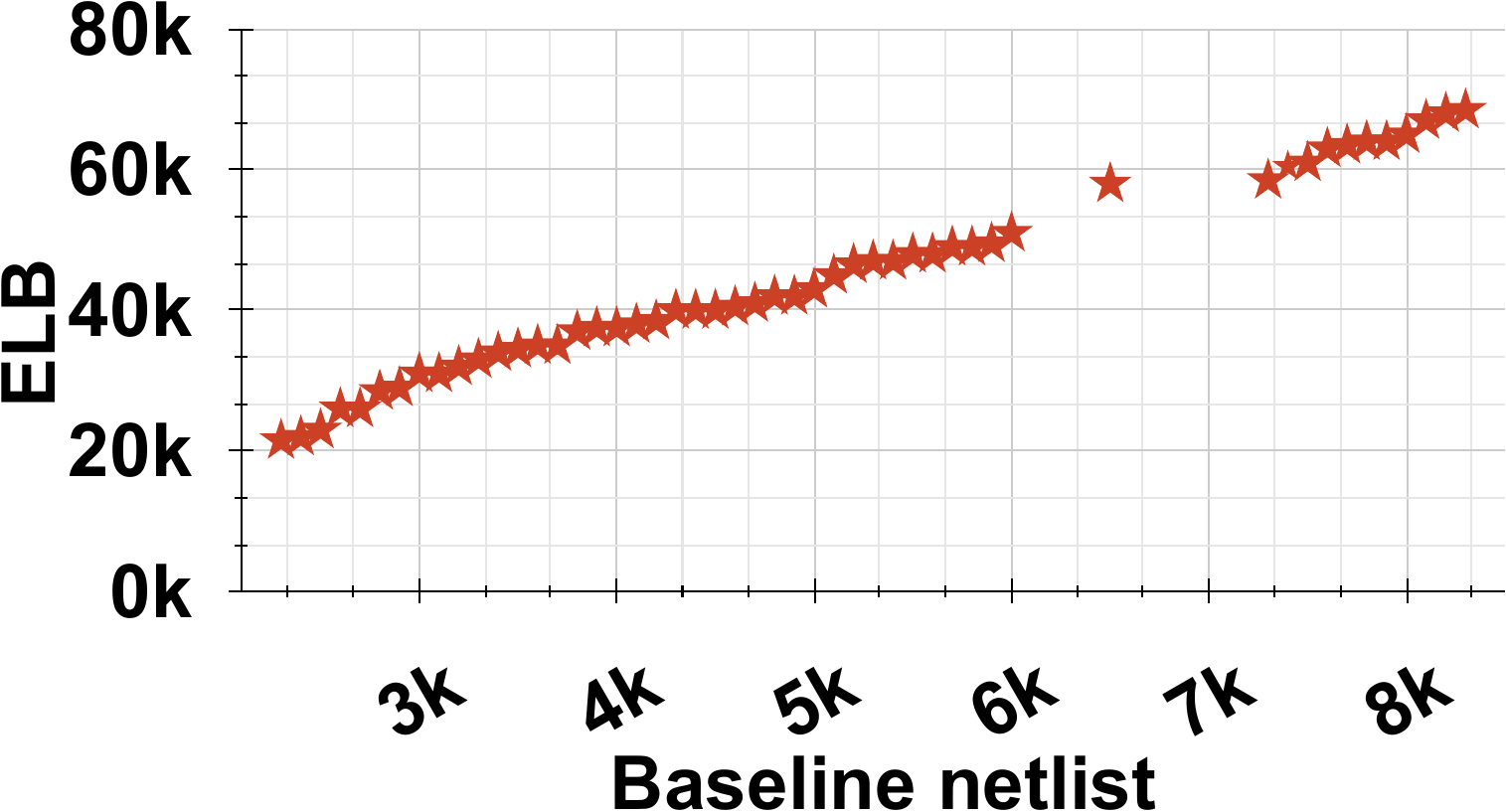}
\caption{$PT_{score}$ on AES before and after applying ELB.}
    \label{fig:elb_motiv}
\end{figure}

We observe a monotonic relationship between pre- and post-countermeasure $PT_{score}$ values.
\autoref{fig:quadseal_motiv}
and \autoref{fig:elb_motiv} show $PT_{score}$ values for 50 randomly synthesized netlists pre- and post-countermeasure application for the two representative countermeasures of choice.\footnote{%
The gaps between 6k and 7k for the baseline/pre-countermeasure netlists
are only due to chance, i.e., none of the 50 random recipes provide scores in that ranges.}

Importantly, this confirms that \ul{aiming for pre-countermeasure resilience suffices for a guided design-space exploration toward best post-countermeasure resilience}.
This observation helps to significantly limit the runtime cost for obtaining training data.
In fact, running S-PSCA evaluations on pre-countermeasure netlists provides a \texttt{7.2x} speed-up: it takes only 40--50 minutes as compared to around 6 hours for post-countermasure netlists.
This is due to the lower resilience of the pre-countermeasure netlists; S-PSCAs can find the correct key with less traces and in shorter time.


\subsubsection{Dataset Diversity}
We are inspired by the sampling approaches described in \cite{bai2023towards}.
These works inspire us to focus on exploring netlists that result in diverse scores.
Without loss of generality, we utilize \ac{SA} toward this end.
Note that we tune the annealing scheduling for more diversity during exploration.

\subsubsection{Training of the Predictor}
Our predictor is a regressor model.
It uses the pre-countermeasure netlists, three handcrafted features,
and $PT_{score}$ values (obtained by actual S-PSCA evaluation) as labels.
The features are:
(1)~\textit{Overall Diversity:} count of the various cell types found in the synthesized netlists;
(2, 3)~\textit{Specific Diversity:} percentage of area consumed by LVT (2) and HVT (3) cells.
We trained and evaluated various predictor versions, exploring a wide range of other hand-crafted features (e.g., area, delay, etc.) as well as more synthesized design and more varied recipes.
However, the best performance, as in the best prediction of post-countermeasure resilience based on pre-countermeasure resilience, was obtained for the three features above.

\subsection{Monte-Carlo Tree Search \& Fine-Tuning (\circletwo{})}

Having established a fast and accurate zero-shot predictor, we now have to integrate it into an search strategy for maximizing PSCA resilience.
\Acf{MCTS}, with its ability to intelligently balance exploration and exploitation, is a natural fit for this complex task.
Its delayed reward model aligns with our problem.



\subsubsection{MCTS Implementation}

We employ MCTS as outlined in Sections~\ref{sec:bg:MCTS} and \ref{sec:problem}.
More details are provided next.

We implement the critical reward component as follows.
We assign normalized $PT_{score} (S_T)$ values, obtained from the zero-shot predictor $PT_{score}$, to a terminal state $S_T$.
\begin{equation}
    PT^{norm}_{score}(S_T)=
    \begin{dcases}\label{eq:1}
        0 & \text{if } PT_{score} \leq PT_{threshold}\\
        \frac{PT_{score}}{PT_{threshold}} & \text{otherwise}
    \end{dcases}
\end{equation}
In plain words,
we compare the predicted scores against a user-defined threshold ($PT_{threshold}$) and assign $0$ reward if it is less than the threshold.
This reward formulation allows \ac{MCTS} to skip any unpromising paths in the search space.
The reward for promising paths are normalized so that the UCT
computation maintains the balance for exploitation and exploration.

For the expansion and rollout stages, note the following.
A sequence of actions are taken which in turn synthesize the netlist until
the terminal state is reached. Then, the $\hat{PT}_{score}$ predictor is used to obtain $PT_{score}$ values of the netlist.
A new node is added and assigned the updated $R(S_t,a_t)$ vale.

\subsubsection{Online Fine-Tuning of Predictor}
As indicated, we simulate the actual power traces for the top-$k$ and bottom-$k$ recipes during the \ac{MCTS} process and accordingly fine-tune the predictor $\hat{PT}_{score}$.
This is done to ensure the predictor is continually updated with relevant corner-case data points, all without hampering the ongoing search-space exploration.
For efficiency,
   we conduct the S-PSCA evaluation for these netlist in parallel.


\subsubsection{Justification of MCTS}

We note that reinforcement learning (RL)-based methods like~\cite{hosny2020drills,zhu2020exploring} are gaining significant attention for advancing logic synthesis.
Important key differences for our work over these works
are as follows, making MCTS uniquely suitable.

In RL-based methods, the
asynchronous actor-critic approach encounters challenges in delayed reward systems.
Thus, RL-based methods tuned for PPA optimization typically assign immediate rewards, like reduced depth of AIGs, which also
correlate well with the final PPA. However, there is no such direct correlation between AIGs and S-PSCA resilience.
Therefore, we cannot utilize immediate rewards and, thus, chose
MCTS. In fact, its notion of {back-propagation} (Section~\ref{sec:bg:MCTS}) helps to accurately estimate average rewards even for intermediate nodes in the search tree.


\subsection{Integration and End-to-End Validation (\circlethree{})}

After completion of the MCTS process, we obtain the best synthesis recipe using a greedy approach following standard procedures
(
most-rewarding-node selection and most-visited-node selection).
This recipe is
expected to generate the most S-PSCA resilient design post-countermeasure application.
Finally, we synthesize the circuit using this recipe, apply the countermeasure of choice, and run an actual S-PSCA evaluation to report the
final $PT_{score}$.





\section{Experiments\label{sec:exp}}




\subsection{Setup}

\textbf{AES Implementation.}
We use non-masked Electronic Code Book (ECB) mode AES with S-Box as a look-up table. ECB mode pads data until the length of the block which is 128.

We use a commercial 28nm technology library, focusing on the TT corner (25 degrees Celsius, 0.9V), and using different VT cells.

\textbf{S-PSCA Setup.}
We implement a C++-based CPA attack following \cite{knechtel22_PSC}. It incrementally increases the number of traces through coarse and thorough sampling for 128 trials, aiming for a 90\% success rate with thorough sampling. In other words, the attack is thoroughly assessed in multiple runs, not only one-shot trials.

The total number of traces depends on the case study; for baseline AES, up to 10K traces are sufficient, whereas QuadSeal and ELB countermeasures on top require 50K and 100K traces, respectively.

\textbf{$PT_{score}$ Predictor.}
We collected a corpus of 1,000 random, different synthesis recipes for the baseline AES, with features extracted as in Section~\ref{sec:method}. This took us $\approx$30 days. We split the dataset into 80\% and 20\% for training and testing, respectively.

We used XGBoost~\cite{chen2016xgboost}, a scalable and distributed, gradient-boosted decision tree (GBDT) library to implement the predictor model.
We trained our model on an \textit{AMD EPYC 7542} server equipped with 128 CPUs and 1TB RAM, running Red Hat Enterprise Linux Server Release 7.9 (Maipo). The CAD flow utilized both open-source and commercial tools, including \textit{Synopsys VCS M-2017.03-SP1} for RTL and gate-level simulations, \textit{Yosys} for logic synthesis, and \textit{Synopsys PrimeTime PX M-2017.06} for power simulations.


\textbf{ASCENT Framework.}
We developed \solution{} in-house by implementing MCTS algorithm and plugged it with $PT_{score}$ . We compute the handcrafted features generated from the synthesized netlist and pass it to $PT_{score}$ which provides quick feedback with a latency of 100 ms. We use Yosys (v0.38) to perform end-to-end synthesis which takes roughly $\sim$2.8 minutes on a single thread CPU run on our Intel server (Frequency: 2.3GHz, Memory: 256GB).

\subsection{Results}

\subsubsection{$PT_{score}$ Predictor}

\begin{figure}[t]
\centering
\includegraphics[width=\columnwidth]{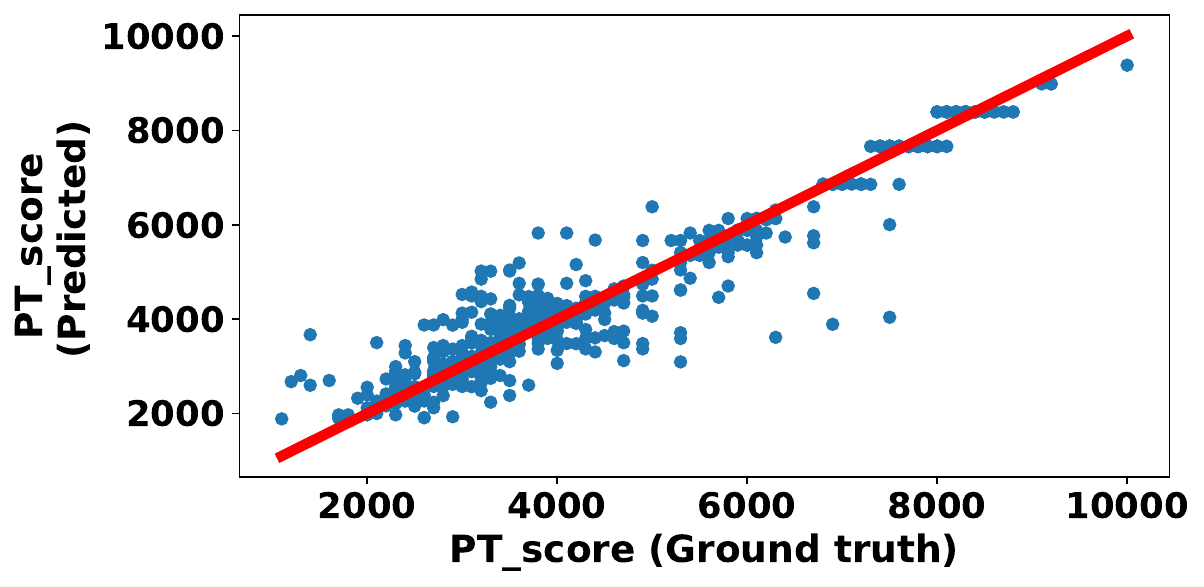}
\caption{Performance of $\hat{PT}_{score}$ predictor on test data.}
    \label{fig:evaluator}
\end{figure}

\autoref{fig:evaluator} shows our predictor's performance on the test datasets, where the x-axis and y-axis denote ``Ground Truth'' and ``Predicted Scores'', respectively.  The Root Mean Squared Error (RMSE) score for the model on the test datasets is \textbf{565.58}. However, recall that the predictor will be continuously improved by fine-tuning to further enhance its performance.

\subsubsection{\solution{} Framework}

\begin{table*}[tbh]
\caption{
	Comparison of $PT_{score}$ across recipes. Shown are only recipes with maximal $PT_{score}$.
}
\begin{center}
\begin{tabular}{|l|c|c|c|c|c|c|c|c|c|c|c|c|}
\hline
\multirow{3}{*}{Synthesis recipes} & \multicolumn{4}{c|}{Baseline AES} & \multicolumn{4}{c|}{AES + QuadSeal} & \multicolumn{4}{c|}{AES + ELB} \\ \cline{2-13}
& PT & Area & Power & Delay & PT & Area & Power & Delay & PT & Area & Power & Delay \\ 
& score & ($\mu$m$^2$) & (mW) & (ns) & score & ($\mu$m$^2$) & (mW) & (ns) & score & ($\mu$m$^2$) & (mW) & (ns) \\ \hline

compress2rs & 2800 & 12391 & 0.65 & 0.28 & 12500 & 16430 & 0.83 & 0.31 & 28000 & 24321 & 1.31 & 0.36 \\

resyn2rs & 2900 & 12380 & 0.62 & 0.3 & 13000 & 16712 & 0.86 & 0.33 & 29000 & 24627 & 1.39 & 0.39 \\

SA & 4900 & 12542 & 0.73 & 0.31 & 17500 & 17123 & 0.91 & 0.34 & 42000 & 25242 & 1.48 & 0.41 \\

MCTS & 5300 & 13723 & 0.76 & 0.31 & 18500 & 18204 & 0.93 & 0.35 & 47000 & 26111 & 1.54 & 0.43\\

\textbf{ASCENT} & \textbf{9100} & \textbf{13155} & \textbf{0.74} & \textbf{0.3} & \textbf{42500} & \textbf{18097} & \textbf{0.92} & \textbf{0.35} & \textbf{87000} & \textbf{25928} & \textbf{1.53} & \textbf{0.44}\\

\hline
\end{tabular}

\label{tab:result1}
\end{center}
\end{table*}

We started with running the S-PSCA evaluation for the \texttt{compress2rs} recipe, which we are considering as the baseline. This will be compared against for all the methods used in this work. The recipe for \texttt{compress2rs} is provided in~\cite{github}.
We consider three different settings for our experiments, namely \textbf{\{AES, AES+QuadSeal, AES+ELB\}}, where AES denotes the baseline without countermeasures, AES+QuadSeal denotes AES with the countermeasure QuadSeal, and AES+ELB represents AES with countermeasure ELB incorporated (Section~\ref{sec:counter}).

\autoref{tab:result1} details our results and comparison across the various methods studied in this work. 
To better understand the results,
we have compared the $PT_{score}$ and overheads with the results from the baseline \texttt{compress2rs} run.
For example for the baseline AES, we would compare the results of each method with the corresponding values from \texttt{compress2rs}.
Here we observed \textbf{\{1.04$\times$, 1.04$\times$, 1.04$\times$\}} improvement in the $PT_{score}$ values, along with an additional \textbf{\{-0.09\%, 1.7\%, 1.26\%\}} area, \textbf{\{-4.6\%, 3.61\%, 6.11\%\}} power, and \textbf{\{7.14\%, 6.45\%, 8.33\%\}} delay overhead across those 3 different scenarios, respectively.
Thus, with the introduction of the countermeasures, we can see that the $PT_{score}$ is improved; this is expected.

 We then utilize the same black-box optimizer that was used for our predictor model, SA, to also explore the security-first search space. Allowing a timeout of 3 days, there is an improvement across all three scenarios: \textbf{\{1.75$\times$, 1.4$\times$, 1.5$\times$\}} higher scores, albeit at an overhead of \textbf{\{1.22\%, 4.22\%, 3.79\%\}} area, \textbf{\{12.31\%, 9.64\%, 12.98\%\}} power, and \textbf{\{10.71\%, 9.68\%, 13.89\%\}} delay, respectively.
 
We then utilized our ASCENT framework to obtain the synthesis recipes with maximized $PT_{score}$ values. Here, an increase in resilience of \textbf{\{1.89$\times$, 1.48$\times$, 1.68$\times$\}}, along with \textbf{\{10.75\%, 10.78\%, 7.36\%\}} area, \textbf{\{16.92\%, 12.05\%, 17.56\%\}} power, and \textbf{\{10.71\%, 12.90\%, 19.44\%\}} delay overheads, respectively, are obtained.

\begin{table}[tbh]
\caption{
	Runtime performance achieved by \solution{} and competitive methods.
 Iterations denote the number of times the full S-PSCA evaluation can be run.
We set a common timeout of 72 hours for fair comparison.
}
\begin{center}
\begin{tabular}{|l|c|r|}
\hline
Method & Iterations & Speed-up \\ \hline
SA (AES with countermeasure) & 12 & 1.0$\times$\\
MCTS (AES with countermeasure) & 12 & 1.0$\times$\\
SA (baseline AES) & 108 & 9.0$\times$\\
MCTS (baseline AES) & 107 & 8.3$\times$\\
\textbf{ASCENT} & \textbf{1460} & \textbf{121.3$\times$}\\ \hline
\end{tabular}

\label{tab:result2}
\end{center}
\end{table}

Finally, we used our full ASCENT framework to more effectively explore the design space, again with timeout of 3 days.
Recall that, thanks to the predictor model,
the time-consuming actual S-PSCA evaluation can be bypassed in the MCTS exploration phase.
We pick the top-3 recipes in terms of max $PT_{score}$ and run actual S-PSCA evaluation on them, for final validation.
From Table 1, it can be seen that, relative to the baseline, we achieved on average across these top-3 recipes \textbf{\{3.25$\times$, 3.40$\times$, 3.11$\times$\}} higher PSC resilience with an overhead of only \textbf{\{6.16\%, 10.15\%, 6.61\%\}} area, \textbf{\{13.85\%, 10.84\%, 16.79\%\}} power, and \textbf{\{7.14\%, 12.90\%, 22.22\%\}} delay, respectively.

\autoref{tab:result2} reports the runtime comparisons of various methods. The first two cases, \textit{\{ SA (AES with countermeasures), MCTS (AES with countermeasures) \}} require running the S-PSCA evaluation end-to-end, with countermeasures in place.
These required to collect 50k and 100k traces for QuadSeal and ELB, respectively, compared to only 10k traces for the baseline AES. Thus, 5$\times$ and 10$\times$ more time is required, respectively.
For the remaining two cases, namely the baseline AES for SA versus MCTS, we obtained a speedup of 9.0$\times$ and 8.3$\times$, respectively.
In short, ASCENT is able to explore a much larger design space much more quickly when compared to other methods; e.g., 120$\times$ faster than the default black-box optimizer SA. 

\section{Conclusion\label{sec:conclusions}}
In this work, we proposed \solution{}, a novel synthesis framework.
We have successfully enhanced the final resilience of existing power side-channel countermeasures by carefully guided, yet efficient, exploration of the complex search space.
In fact, \solution{} enables a $3.11\times$ post-countermeasure improvement when compared to conventional synthesis (tailored for PPA optimization).
For future work, we plan to tailor \solution{} to harden circuits also against other threats like fault injection.

\printbibliography

\end{document}